\documentclass[prb,twocolumn,floatfix,notitlepage,superscriptaddress,longtable]{revtex4-2}
\usepackage{amsmath}
\usepackage{lipsum} % for some text
\usepackage{verbatim}
\usepackage{epsfig}
\usepackage{subfigure}
\usepackage{graphicx}
\usepackage{amsfonts}
\usepackage[figuresright]{rotating}
\usepackage{amssymb}
\usepackage{amsmath}
\usepackage{psfrag}
\usepackage{esint} %double integrals
\usepackage{bm}% bold math
\usepackage[colorlinks,linkcolor=blue,anchorcolor=blue,citecolor=blue,urlcolor=blue]{hyperref}
\usepackage[version=4]{mhchem}
\usepackage[svgnames]{xcolor}
\def\be{\begin{equation}} \def\ee{\end{equation}}
\def\bea{\begin{eqnarray}} \def\eea{\end{eqnarray}}

\def\bk{{\bf k}}

\renewcommand{\vec}[1]{\mathbf{#1}}

\def\bpm{\begin{pmatrix}} \def\epm{\end{pmatrix}}

\DeclareMathOperator{\Tr}{Tr}

\definecolor{Qicolor}{RGB}{3, 136, 252}

\usepackage{framed}
\usepackage{listings}
\usepackage{color}
\usepackage{tikz}
\usetikzlibrary{shapes.arrows}
\usetikzlibrary{backgrounds}
\usetikzlibrary{positioning}
\usetikzlibrary{arrows}
\definecolor{gray}{gray}{0.3}
\definecolor{darkgreen}{rgb}{0,0.55,0}
\definecolor{purple}{rgb}{0.5,0,1}

\makeatletter
\newcommand*{\balancecolsandclearpage}{%
  \close@column@grid
  \clearpage
}
\makeatother

\begin{document}

\author{Omid Tavakol}
\affiliation{Department of Physics and Astronomy, University of California, Irvine, California 92697, USA}
\affiliation{Department of Physics, University of Toronto, 60 St. George Street, Toronto, Ontario, M5S 1A7, Canada}
\author{Thomas Scaffidi}
\email{tscaffid@uci.edu}
\affiliation{Department of Physics and Astronomy, University of California, Irvine, California 92697, USA}
\affiliation{Department of Physics, University of Toronto, 60 St. George Street, Toronto, Ontario, M5S 1A7, Canada}

\title{Minimal model for the flat bands in copper-substituted lead phosphate apatite: \\ 
Strong diamagnetism from multi-orbital physics}

\begin{abstract}
The claims that a copper-substituted lead apatite, denoted as CuPb$_9$(PO$_4$)$_6$OH$_2$, could be a room-temperature superconductor have led to an intense research activity. 
While other research groups did not confirm these claims, and the hope of realizing superconductivity in this compound has all but vanished, other findings have emerged which motivate further work on this material. In fact,
Density Functional Theory (DFT) calculations indicate the presence of two nearly flat bands near the Fermi level, which are known to host strongly correlated physics. 
In order to facilitate the theoretical study of the intriguing physics associated with these two flat bands, we propose a minimal tight-binding model which reproduces their main features. 
We then calculate the orbital magnetic susceptibility of our two-band model and find a large diamagnetic response which arises due to the multi-orbital nature of the bands and which could provide an explanation for the strong diamagnetism reported in experiments.
\end{abstract}
\date{\today}

\maketitle

\emph{Introduction}---
Recently copper-substituted lead apatite CuPb$_9$(PO$_4$)$_6$OH$_2$ was proposed as a room-temperature superconductor~\cite{lee2023roomtemperature,lee2023superconductor}.
While superconductivity in this material was not replicated by other groups, a Density Functional Theory (DFT) calculation recently predicted the appearance of two nearly flat bands in this material close to the Fermi level~\cite{Griffin} (see also Refs.~\cite{DFT1,DFT2,DFT3,DFT4}). 
Since flat bands are known to host strongly correlated physics~\cite{Heikkila2011,Kopnin_2011,Hofmann_2020,Torma2022}, this motivates additional research on CuPb$_9$(PO$_4$)$_6$OH$_2$. Besides, the reported unusual magnetic properties of this material--- a combination of soft ferromagnetism and strong diamagnetism~\cite{FMDM}--- provide further motivation.
In order to study the potentially rich physics hosted by the two flat bands, it would be beneficial to have a minimal two-band model which reproduces their main qualitative features.
In this letter, we propose such a minimal model, and discuss the implications for superconductivity and diamagnetism.

\begin{figure}[t!]
    \centering
    \includegraphics[width=0.25\textwidth]{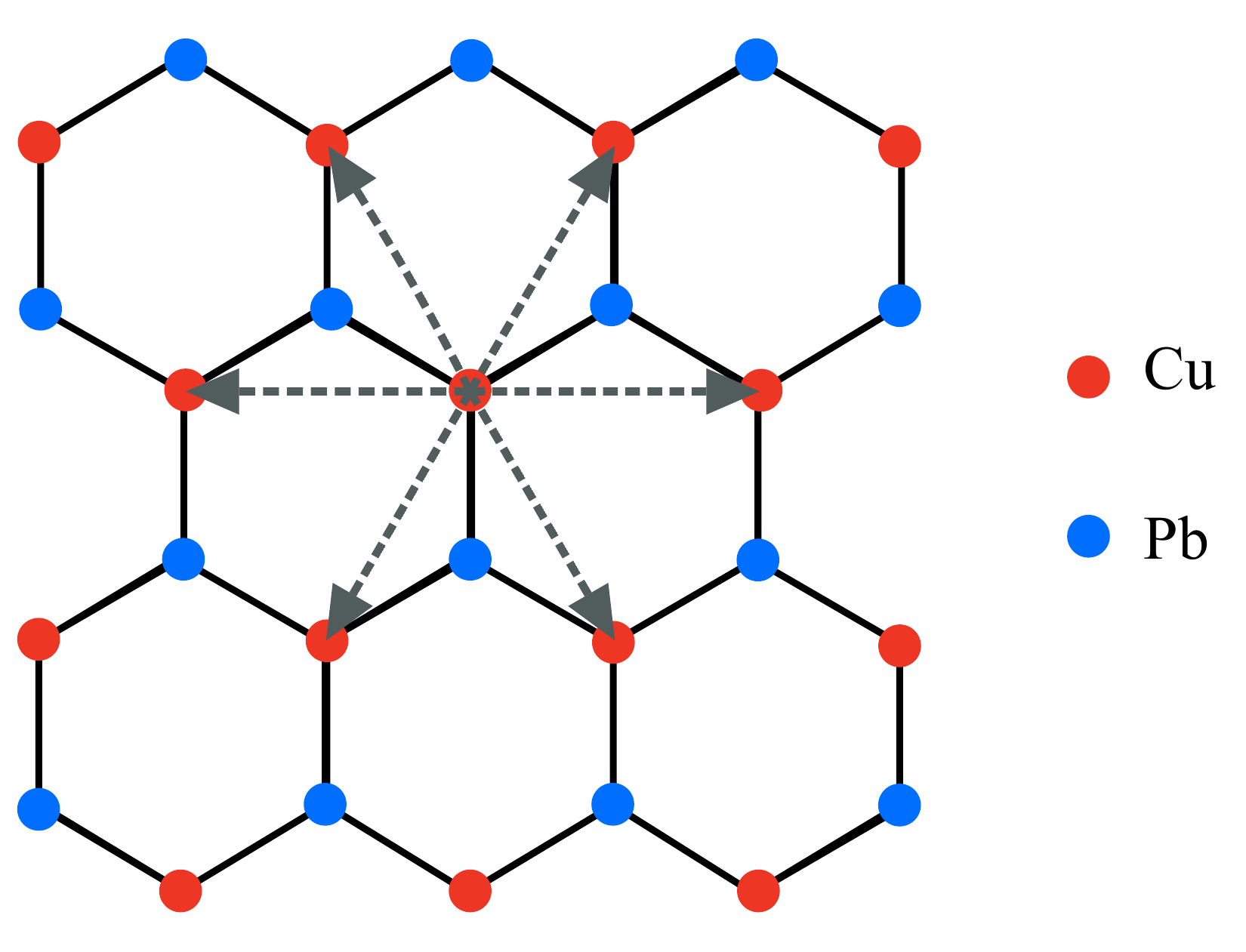}
    \caption{Pb(1) sites forming a hexagonal lattice in the lead apatite structure. After substitution by copper, one sublattice is occupied by Cu (which we call the A sublattice, in red) and the other one is occupied by Pb (which we call the $B$ sublattice, in blue).\label{fig:lattice}}
\end{figure}

\emph{Model}---
In the apatite structure, the Pb(1) sites form a honeycomb lattice. (We follow here the nomenclature in Ref.~\cite{Griffin} for Pb(1) and Pb(2) sites but we note that the opposite naming convention also appears in the literature).
Following Ref.~\cite{Griffin}, we investigate the situation in which Pb is substituted by Cu on one of the two sublattices of the Pb(1) sites (see Fig.~\ref{fig:lattice}), and in which the Cu substition sites are stacked vertically on every other layer.
In that case, Ref.~\cite{Griffin} predicts the existence close to the Fermi level of two isolated ``flat'' bands with a small bandwidth (around 0.1 eV), which are well-separated from all the other bands.
Since these two bands are predicted to be mostly of Cu $d_{zx}$ and $d_{zy}$ character, we aim to write a minimal tight-binding two-orbital model based on these orbitals located on the A sublattice of the honeycomb Pb(1) lattice.
Even though they have a small bandwidth, these two bands still have a non-zero momentum dependence (See Fig 4 of Ref.~\cite{Griffin}) which should be important for the physics, and which we hope to reproduce with our model.

We propose the following single-particle Hamiltonian $H_0 = \sum_{\mathbf{k}} c^\dagger_{\bk,a} H_{a,b}(\bk) c_{\bk,b}$ with: 
\bea
H_{a,b}(\mathbf{k}) =  d_\nu (\mathbf{k})  \sigma^{\nu}_{a,b} - \mu  \delta_{a,b}
\label{Eq:Hamiltonian}
\eea
with $\sigma^{\nu} = ( \mathbf{1}, \sigma^x, \sigma^y, \sigma^z)$ the Pauli matrices written in the orbital basis indexed by $a,b=\{d_{zx},d_{zy}\}$, with $\mu$ is the chemical potential, and with
\bea
d_0(k) &=& -\frac 12 \left( \frac12 \sum_{i=1,2,3} \cos(\mathbf{k} \cdot \mathbf{b}_i) +3 \right) - 2 t_z \cos(k_z) \\
d_1(k) &=& -\frac{\sqrt{3}}{4} ( -\cos( \mathbf{k} \cdot \mathbf{b}_1) + \cos(\mathbf{k} \cdot \mathbf{b}_2) ) \nonumber\\
d_2(k) &=& -\frac{\sqrt{3}}{4}  \left( \sin(\mathbf{k} \cdot \mathbf{b}_3) +  \sin(\mathbf{k} \cdot \mathbf{b}_2) +  \sin(\mathbf{k} \cdot \mathbf{b}_1) \right)  \nonumber \\
d_3(k) &=& -\frac12 \left( \cos(\mathbf{k} \cdot \mathbf{b}_3) - \frac12  (\cos(\mathbf{k} \cdot \mathbf{b}_1) + \cos(\mathbf{k} \cdot \mathbf{b}_2) ) \right)\nonumber 
\eea
where we used the lattice vectors connecting nearest-neighbor Cu sites: $\mathbf{b}_1 = - \sqrt{3}/2 \hat{e}_x + 3/2 \hat{e}_y$, $\mathbf{b}_2 =  - \sqrt{3}/2 \hat{e}_x - 3/2 \hat{e}_y$, and $\mathbf{b}_3 =  \sqrt{3} \hat{e}_x$.
In real space, this Hamiltonian reads
\bea
H_0  &=& \sum_{\mathbf{r}} \left(-\mu -\frac32 \right) c^\dagger_{\mathbf{r},a} c_{\mathbf{r},a}\nonumber\\
&-&\sum_{\mathbf{r}} \sum_{i=1,2,3} c^\dagger_{\mathbf{r},a}  (\mathcal{T}_i)_{a,b}   c_{\mathbf{r}+\mathbf{b}_i,b} + \text{h.c} \nonumber \\
&-& t_z \sum_{\mathbf{r}} c^\dagger_{\mathbf{r},a} c_{\mathbf{r}+c\hat{z},a} + \text{h.c} 
\eea
where $\mathbf{r}$ is summed over Cu sites, $c$ is the lattice spacing between copper sites along the $z$ axis, and where the hopping matrices are given by
\bea
\mathcal{T}_1 = \begin{pmatrix}
0 & -\frac{\sqrt{3}}{4} \\
0  & \frac{1}4 
\end{pmatrix},
\eea
and
\bea
\mathcal{T}_2 = \begin{pmatrix}
0 & 0 \\
\frac{\sqrt{3}}{4}  & \frac{1}4 
\end{pmatrix},
\eea
and
\bea
\mathcal{T}_3 = \begin{pmatrix}
\frac38 & -\frac{\sqrt{3}}{8} \\
\frac{\sqrt{3}}{8}  & \frac{-1}8 
\end{pmatrix}.
\eea
The $-3/2$ potential term can of course be absorbed in the chemical potential but we keep it as it makes the connection to the honeycomb $p_x, p_y$ honeycomb model more transparent~\cite{PhysRevLett.99.070401}, as we show below.

\begin{figure}[t!]
    \centering
    \includegraphics[width=0.45\textwidth]{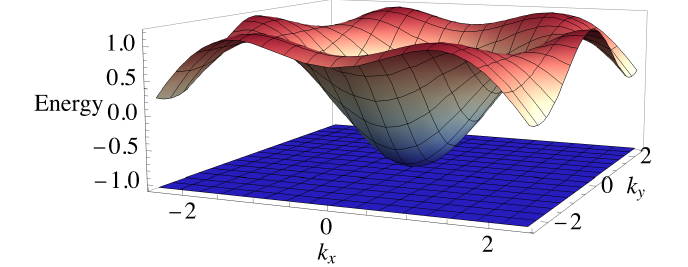}
    \caption{Surface plot of $E_{\pm}(\mathbf{k})$ at $k_z=0$, showing the bottom flat band and the top dispersive band.\label{fig:surfaceplot}}
\end{figure}

This Hamiltonian contains hopping terms to the 6 nearest-neighbors on the triangular lattice of the Copper sites (see arrows in Fig.~\ref{fig:lattice}), and to the two nearest-neighbors along the $z$ axis (the Pb(1) sites are stacked on top of each other along the $z$ axis). 
The orbital character of the planar hoppings reflects the anisotropy of the $d$ orbitals.
We calculated it by starting from a four-band honeycomb lattice model~\cite{PhysRevLett.99.070401} with a $d_{zx}$ and $d_{zy}$ orbital on every site, and for which only nearest-neighbor hopping for orbitals which are parallel to the honeycomb bond is considered since it is dominant.
We then added a strong staggered potential $\epsilon$ which gaps out the $B$ sublattice, leaving behing a two-band model for the $A$ sublattice.
(Such a staggered potential in the four-band model was studied in Ref.~\cite{Hao_2022}).
At leading order in perturbation theory in $1/\epsilon$, we obtain the tight-binding Hamiltonian given in Eq.~1.
Our derivation of the tight-binding model is phenomological in nature and a more microscopic derivation which accounts for the complex environment of the Cu atoms is left for future work.

The dispersion relation is given by $
E_\pm(\mathbf{k}) = d_0(\mathbf{k}) \pm \sqrt{|\vec{d}(\mathbf{k})|^2}
$ and is shown in Fig.~\ref{fig:surfaceplot}.
The bottom band is dispersionless along the planar $k_x,k_y$ directions, with energy $E_- = - 9/4 - 2 t_z \cos(k_z)$.
The top band is dispersive, with energy $E_+ = -\frac14 (3 + 2 \sum_{i=1,2,3} \cos(\mathbf{k} \cdot \mathbf{b}_i) ) - 2 t_z \cos(k_z)$.
Our Hamiltonian is defined in arbitrary units and should be scaled by roughly 0.05 eV in order to reproduce the bandwidth observed in Fig.~\cite{Griffin}.

\begin{figure}[t!]
    \centering
    \includegraphics[width=0.45\textwidth]{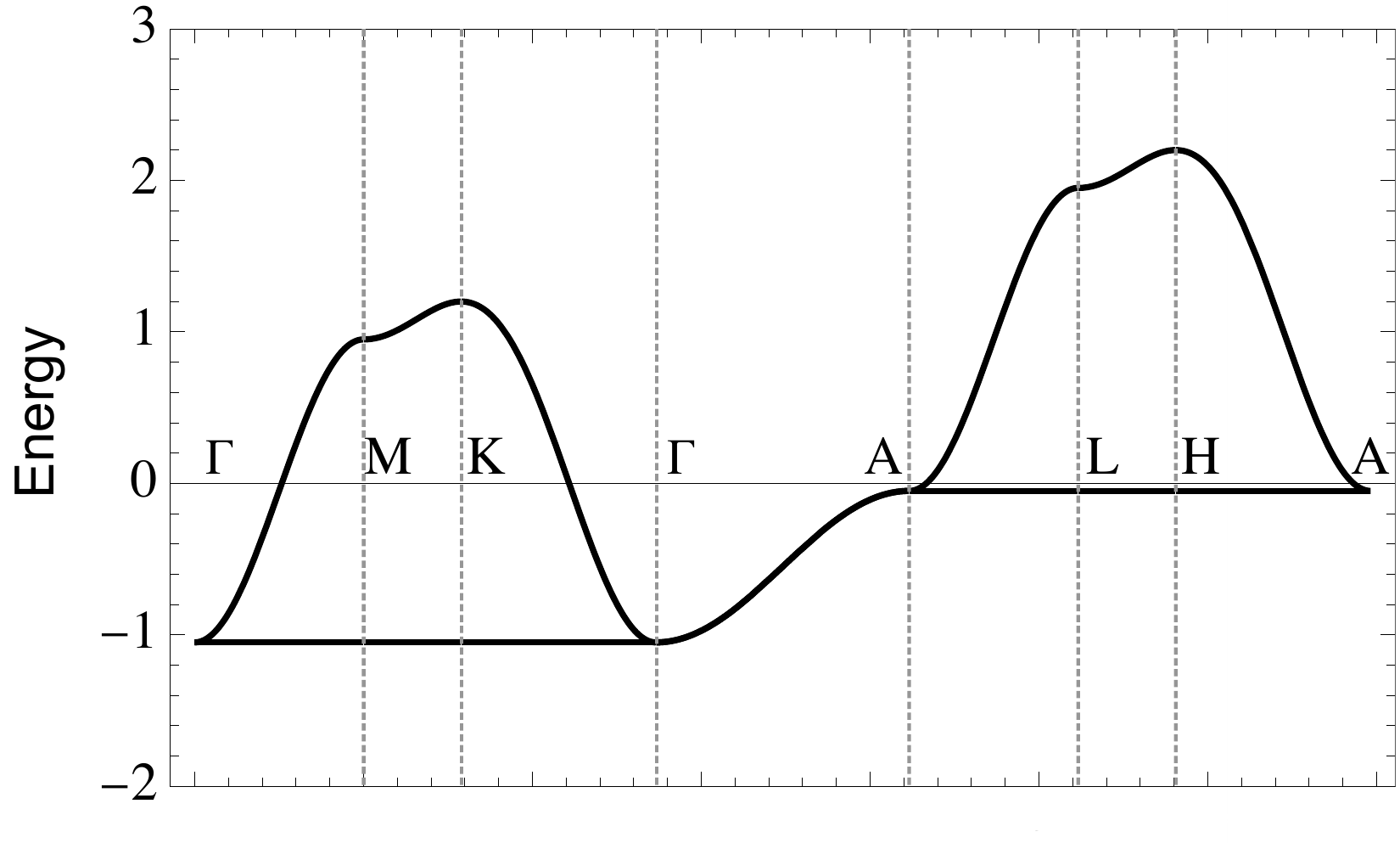}
    \caption{Dispersion relation $E_{\pm}(\mathbf{k}) - \mu$ with $t_z = 0.25$ and $\mu=-1.7$. The points A, L, and H are all at $k_z=\pi$. \label{fig:bandplot}}
\end{figure}
As shown in Fig.~\ref{fig:bandplot}, our model reproduces the main qualitative features of the two flat bands (see Fig 4 of Ref.\cite{Griffin}): a top band whose planar dispersion is much larger than the bottom band, a quadratic band touching at the Gamma point, and a simple $\cos(k_z)$-like dispersion for both bands. (In our case, the bottom band is perfectly dispersionless in the plane, but adding further hopping terms would make it possible to change that and reproduce the small dispersion seen in Ref.~\cite{Griffin}).
We chose a value of the chemical potential $\mu=-1.7$ which reproduces the fact that the top band is roughly half-filled at $k_z=0$.
Further, we found that a value of $t_z=0.25$ reproduces the feature that the bottom band is very close to the Fermi level at $k_z=\pi$.

The fact that the bottom band is perfectly dispersionless in our model arises due to the presence of localized states on the three $A$ sites around each hexagon.
This is a triangular lattice version of the flat bands arising in the $p_x,p_y$ model on the honeycomb lattice~\cite{PhysRevLett.99.070401}.
In fact, for $t_z=0$, the bands we find have energies which are the square of the bands found in the $p_x, p_y$ honeycomb model~\cite{PhysRevLett.99.070401}.
Indeed, the $p_x, p_y$ nearest-neighbor honeycomb model has four bands with dispersion $E_{1,4} = \pm \sqrt{9/4}$ and $E_{2,3} = \pm \sqrt{\frac14 (3 + 2 \sum_{i=1,2,3} \cos(\mathbf{k} \cdot \mathbf{b}_i) )}$. 
This is not surprising since our tight-binding model can be obtained starting from the four band model of Ref.~\cite{PhysRevLett.99.070401}, as explained above.

We note that our model is not the same as the $p_x, p_y$ model on a triangular lattice considered in Ref.~\cite{trianglepxpy}. 
The reason is that, in our case, the underlying honeycomb lattice breaks inversion symmetry and changes the orbital character of the dominant hopping terms.

It is noteworthy that the dispersive band  $E_+$ has the same dispersion relation as a single-orbital nearest-neighbor model on a triangular lattice (see Fig.~\ref{fig:surfaceplot}). However, this simple dispersion relation actually hides a non-trivial $k$-dependence of the orbital character of the bands which is of course absent for the single-orbital triangular lattice model.
We now discuss the possible implications of this orbital character of the bands for superconductivity and for the orbital magnetic susceptibility.

\emph{Superconductivity}---
% Flat bands can host a variety of strongly correlated phases, including magnetic phases and charge-density waves~\cite{Derzhko2015,Hofmann_2023,Hase2023}, but we will discuss first superconductivity given the recent claims made in Refs.~\cite{lee2023roomtemperature,lee2023superconductor}.
Even though the claims of superconductivity in this compound were not confirmed, it is still interesting to study theoretically how the orbital character of the bands in our model constrain the possible superconducting order parameters.
Assuming a phonon mechanism for superconductivity, it is natural to consider on-site pairing.
We note that the DFT calculations of Ref.~\cite{Griffin} predict that the flat bands are perfectly spin-polarized.
Assuming that this is indeed the case, the only on-site pairing term allowed by the Pauli principle is an orbital singlet:
\bea
H_{SC} = \Delta \sum_{\mathbf{r}} (c^\dagger_{\mathbf{r},zx} c^\dagger_{\mathbf{r},zy}  - c^\dagger_{\mathbf{r},zy} c^\dagger_{\mathbf{r},zx}) + \text{h.c.}
\eea

\begin{figure}[t!]
    \centering
    \includegraphics[width=0.35\textwidth]{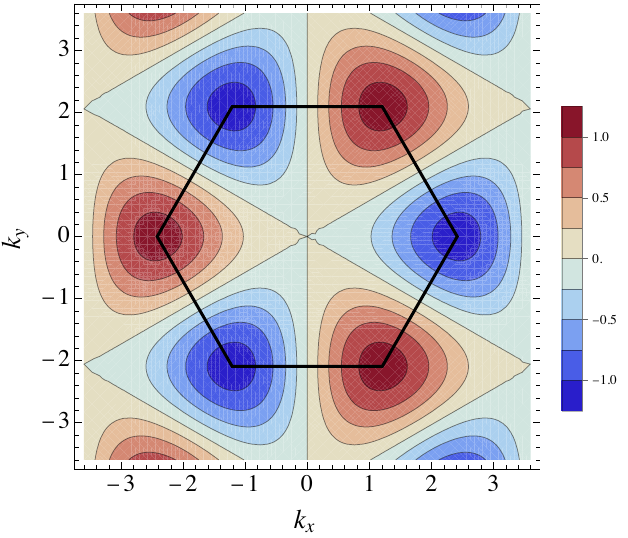}
    \caption{Contour plot of $\Delta(\bk) \propto d_2(\bk)$ for on-site pairing, showing an f-wave pattern. The Brillouin zone is shown in black.  \label{fig:contourplot}}
\end{figure}

Since at $k_z=0$, only the dispersive top band crosses the Fermi level, let us keep only intraband pairing terms on the top band, leading to
\bea
H_{SC} = \sum_{\mathbf{k}} \Delta(\mathbf{k}) c^\dagger_\mathbf{k} c^\dagger_{-\mathbf{k}} + \text{h.c.}
\eea
with
\bea
\Delta(\mathbf{k}) = \Delta (u_x^*(\mathbf{k}) u_y^*(-\mathbf{k}) - u_y^*(\mathbf{k}) u_x^*(-\mathbf{k}) )
\eea
where $(u_{x}(\mathbf{k}), u_{y}(\mathbf{k})) $ is the Bloch eigenvector of the top band, and where we kept the band index implicit.
Using the parametrization $\mathbf{d} = (\sin(\theta/2) \cos(\phi), \sin(\theta/2) \sin(\phi) , \cos(\theta/2))$, we have $(u_x,u_y) = (\cos(\theta/2), e^{i \phi} \sin(\theta/2))$.
Using the fact that $\theta(\bk) = \theta(-\bk)$ and $\phi(\bk) = - \phi(-\bk)$, one finds $\Delta(\bk) \propto d_2(\bk)$.
As seen in Fig.~\ref{fig:contourplot}, on-site pairing thus generates an f-wave gap of the type $\Delta(\bk) \sim k_x (k_x^2 - 3 k_y^2)$.
We should emphasize that this f-wave gap is on-site and is thus microscopically distinct from the non-on-site f-wave gaps which are known to occur in the weak coupling limit of the triangular lattice single-orbital Hubbard model~\cite{PhysRevB.81.224505}.
We also note that inter-orbital pairing in non-centrosymmetric systems was also discussed in the context of other materials~\cite{Tanaka}, including transition metal dichalocogenides~\cite{MARGALIT2021168561}.

Assuming an electronic mechanism for superconductivity, one could start from the on-site repulsive inter-orbital interaction term (which is the only on-site interaction term within our model if we again assume perfectly spin-polarized bands)
\bea
H_{int} = U \sum_{\mathbf{r}} n_{\mathbf{r},d_{zx}} n_{\mathbf{r},d_{zy}}
\eea
and calculate the effective interaction in the Cooper channel, which would favor non-on-site pairing.
Since $U>0$, and based on the discussion above, on-site repulsion would disfavor any pairing in the f-wave channel.
One should thus consider non-on-site pairing in other channels.
We note that minimal models featuring $d_{zx}$ and $d_{zy}$ orbitals have been used to study superconductivity in a variety of materials, including pnictides~\cite{PhysRevB.77.220503} and Sr$_2$RuO$_4$~\cite{PhysRevLett.105.136401,PhysRevB.107.014505}.

\begin{figure}[t!]
    \centering
    \includegraphics[width=0.45\textwidth]{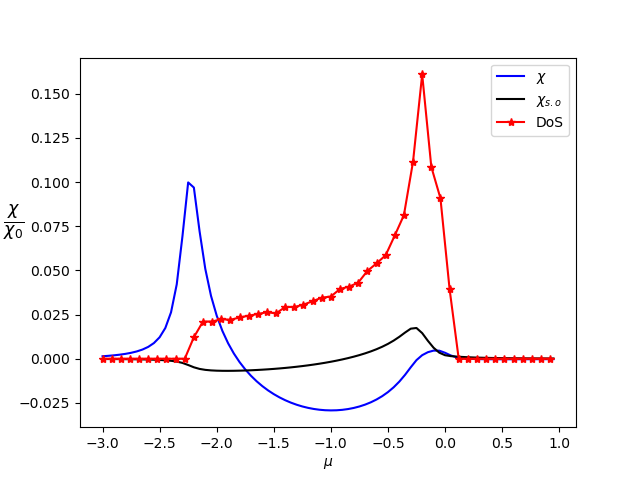}
    \caption{Orbital susceptibility for our two-band model $\chi$  (in blue) and for a single-orbital model $\chi_{s.o}$ (in black). The density of states (DoS) of the top band $E_+$ (which is the same as the the DoS of the single-orbital model) is also shown in red. We have $\chi_0 = \frac{\mu_0 e^2|t| a^2}{\hbar^2}$, with $|t|$ the magnitude of the nearest-neighbor hopping term, and $a$ the lattice spacing between copper sites. The DoS is in units of $\frac{1}{|t| a^2}$. We used $\eta = 0.1$ to calculate the susceptibilities.}
    \label{fig:susc}
\end{figure}

\emph{Magnetic susceptibility}---
The magnetic properties of copper-substituted lead apatite were reported to exhibit a combination of soft ferromagnetism and strong diamagnetism~\cite{FMDM}. Motivated by these findings, we calculate the out-of-plane orbital magnetic susceptibility of the Hamiltonian of Eq.~\ref{Eq:Hamiltonian} using the following formula \cite{Koshino_2007,Santos_2011,Santos_2016}
\bea
&\chi = -\frac{\mu_0 e^2}{2\pi \hbar^2}\mathrm{Im} \int_{-\infty}^{\infty} dE  n_F(E)  \nonumber \\ &\frac{1}{A}\sum_{\vec{k}}\Tr \bigg[\gamma_{x}\hat{G}\gamma_{y}\hat{G}\gamma_x\hat{G}\gamma_y\hat{G}\nonumber \\&
+\frac{1}{2}(\hat{G}\gamma_x\hat{G}\gamma_y+\hat{G}\gamma_y\hat{G}\gamma_x)\hat{G}\frac{\partial \gamma_y}{\partial k_x}\bigg]
\label{Eq:susc}
\eea
where $\hat{G}(E,\vec{k}) = (E - H_\vec{k} + i\eta)^{-1}$, $\gamma_{x,y}=\frac{\partial \hat{H}_{\vec{k}}}{\partial k_{x,y}}$, $\mu_0$ is the vacuum permeability, $\eta$ is a small positive number, $A$ is the sample area, and $n_F (E) = (e^{(E-\mu)/T} + 1)^{-1}$ is the Fermi distribution function. For the rest of the discussion we will work at $T=0$, and we use an effectively two-dimensional by restricting $\bk$ to the $k_z=0$ plane.

One notable advantage of employing Eq. (\ref{Eq:susc}) over the Peierls-Landau orbital susceptibility method \cite{Fukuyama1971, Safran1979, Vignale1991} is that it accounts for contributions originating from the quantum geometry of the Bloch wave functions, which has been shown to play a crucial role in multi-orbital models~\cite{Raoux2014}.

In this context, it is instructive to compare the susceptibility of our model with that of a single-orbital, nearest-neighbor triangular lattice Hamiltonian, with dispersion $E_{s.o} = -\frac{3}{4}-\frac{1}{2}\sum_{i=1}^3 \cos(\mathbf{k} \cdot \mathbf{b}_i)$. As noted before, this single-band Hamiltonian has exactly the same dispersion as the upper band of our two-band model. By comparing these two models, we can thus elucidate the specific impact of multi-orbital physics on the orbital susceptibility.

The results for the susceptibility are shown in Fig.~\ref{fig:susc}. 
The results for the single-orbital model are easily interpreted: at low filling, there is a single pocket around the Gamma point and lattice effects become negligible, leading to a negative value for $\chi_{s.o}$, as it should be in order to recover Landau diamagnetism of free electrons. At larger filling, the system goes through a van Hove singularity, at which the Fermi surface changes topology and at which the DoS diverges. This van Hove singularity produces a paramagnetic peak for $\chi_{s.o.}$, which is a well-known effect~\cite{Vignale1991}.

Interestingly, the susceptibility of our two-band model $\chi$ has very different features.
First, there is a paramagnetic peak at low filling (for $-2.5 < \mu <-2$), which is due to the presence of the bottom flat band $E_-$.
Second, there is a broad region of diamagnetism for $-1.75 < \mu < -0.25$, despite the presence of a van Hove singularity close to $\mu=-0.25$. We note that the single-orbital model is paramagnetic in the range $-1.75 < \mu < -0.25$ due to the van Hove singularity, and the diamagnetism of the two-orbital model thus has a multi-orbital origin. The diamagnetism in this region is comparatively strong: the most negative value reached by $\chi$ for $\mu \simeq -1$ is about four times larger than the one reached by $\chi_{s.o}$ at low filling.
The strong diamagnetism we observe in the two-band model could thus provide an explanation for the one reported in experiments~\cite{FMDM}.

In conclusion, we have proposed a minimal two-band model which reproduces the main features of the Cu $d_{zx}$ and $d_{zy}$ flat bands predicted to appear in a copper-substituted lead apatite in Ref.~\cite{Griffin}.
Using this model, we have found a strong diamagnetic response which could explain the one observed in experiments, and whose origin lies in the multi-orbital nature of the bands.
Looking beyond this material, our two-orbital model provides an interesting toy model to study strongly correlated and multi-orbital physics.

\emph{Note added in print}---
Shortly after our manuscript appeared on arXiv, other tight-binding models capturing the flat bands of copper-substituted lead apatite were proposed~\cite{Hirschmann2023,Lee,Yue2023,2023arXiv230805143J,Mao2023}.

\begin{acknowledgments}
TS gratefully acknowledges discussions with Steven White, Javier Sanchez-Yamagishi and Felipe Gonzalez.
\end{acknowledgments}

\bibliography{main}
\end{document}